# Theoretical predictions for ultrasensitive sensing in three-dimensional stochastic interferometry

Guillaume Graciani [*]
*SAMOVAR, Télécom SudParis, Institut Polytechnique de Paris, 91120 Palaiseau, France*

Marcel Filoche
*Institut Langevin, ESPCI Paris, PSL University, CNRS, 75005 Paris, France*

We show that a random light field can be harnessed for high-precision metrology by introducing specific boundary conditions in the form of Lambertian reflections inside a cavity. We demonstrate a quantifiable and reproducible interferometric response to minute perturbations in wavelength, refractive index, and geometry, predicting high sensitivities consistent with experimental measurements of geometrical deformations down to the picometer scale.

## I. INTRODUCTION

Speckle metrology involves harvesting the seemingly random interference pattern formed by a beam of coherent light going through a complex medium to study the properties either of the light or of its supporting medium: ultrasensitive wavelength determination and spectral analysis [1,2], refractive index sensing [3], small angles and displacements metrology [4], or the well-known dynamic light scattering (DLS) [5] and diffusing wave spectroscopy [6–9] for particle sizing and rheology [10,11] to name a few.

While in most of these applications the very speckle pattern used for metrology is directly generated from the medium under study, we propose here a fundamentally different interferometric approach in which a three-dimensional (3D) field of random waves is generated inside a cavity independently from any sample and then used to measure perturbations acting on it such as geometrical changes or local perturbations of the refractive index. This setup maximizes sensitivity through laser resonance phenomena, at the expense of any spatial information on the perturbation. This paper finds support in our recent realization of such a so-called 3D stochastic interferometer, used as a highly sensitive way to homogeneously probe the geometric and dielectric fluctuations of an optical volume [12,13], detect seismic and acoustic vibrations [13], amplify DLS for particle sizing [14], and perform a marker-free study of protein dynamics in solution [15]. In this paper, we theoretically derive the high sensitivity of this setup to perturbations of the medium, and numerically compute noise floors in excellent agreement with experimental realizations.

[*]Contact author: guillaume.graciani@telecom-sudparis.eu

## II. EXPERIMENTAL 3D STOCHASTIC INTERFEROMETER SETUP

The experimental setup consists of a single-frequency laser beam propagating inside a closed, high-reflectivity Lambertian cavity, and a light detection scheme capable of probing the speckle field within it. Both the input and output are realized through single mode optical fibers fit in the cavity through dedicated holes. The randomness and the high statistical symmetry of the optical field arise from the linear transformation of the input laser field through the deterministic yet extremely complex multiple scattering caused by the quenched disorder of the Lambertian wall structure. This highly complex reflection process results in a Lambertian boundary condition that acts as a feedback leading to the interferometric properties of the field.

According to the Berry hypothesis, the wave field at each point inside the cavity can be considered as a coherent superposition of a large number of plane waves with arbitrary directions and random phases, leading to a 3D speckle that is stationary unless the geometry of the cavity or the dielectric properties of its content fluctuate. Due to the statistical homogeneity and isotropy of the light field, all points in the cavity, all directions, and all polarization states are statistically equivalent regardless of the geometry of the cavity [12].

From previous assumptions, a unique property follows: A local perturbation—for instance of the cavity geometry—can strongly affect the phase of the wave field at any other point inside the cavity. As a consequence, the cavity and the field inside can be used as one single interferometer, in the sense that its response does not depend on the cavity shape, how the light is injected, where perturbations take place, or where the response is measured [12]. When injecting a single-frequency laser at a given input point, each wave field that is observed at any other location is coupled to the laser field at the input point, and its random phase essentially reflects the phase lag corresponding to the length of the random propagation path that connects the two points (see Fig. 1). In the following, we will consider three types of average for any measured quantity $X$: $\overline{X}$ will denote a spatial average, $\langle X \rangle_t$ a time average, and $\langle X \rangle_{\mathcal{C}}$ an average over all speckle configurations.

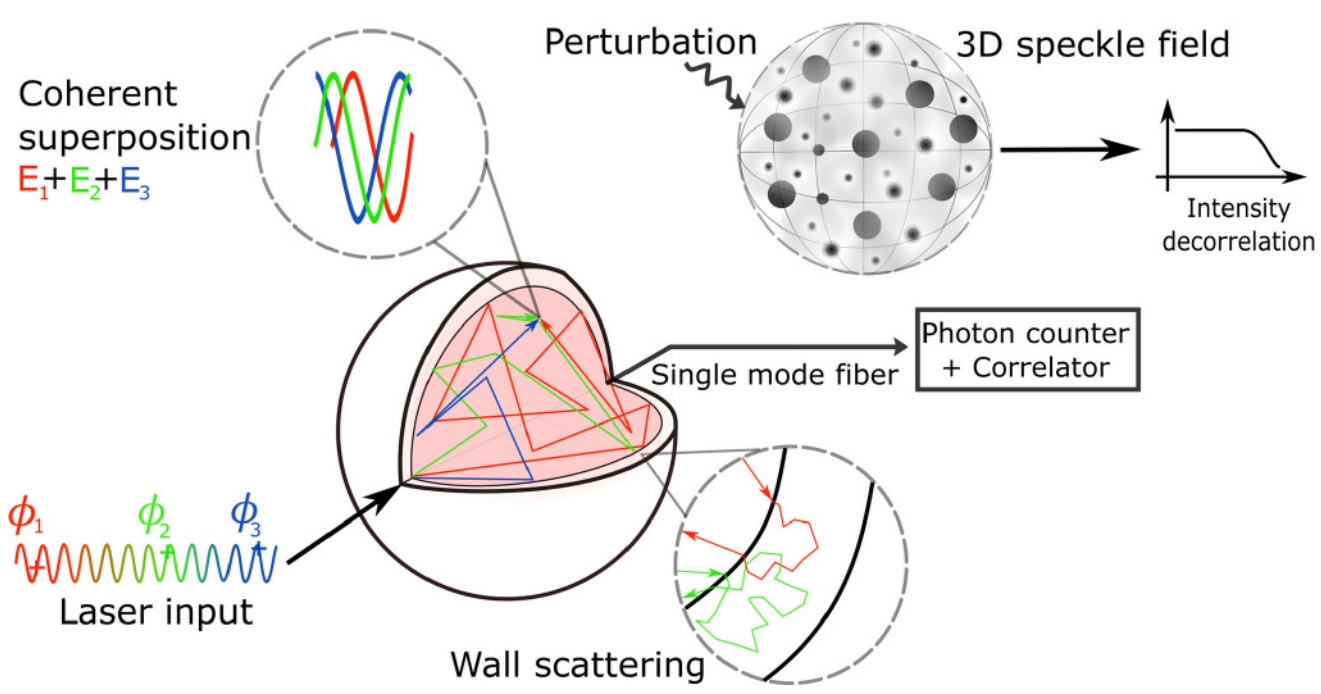

FIG. 1. Schematics of the setup: A monochromatic laser is injected inside a Lambertian cavity. Three light paths of different lengths are represented in blue, green, and red (from shortest to longest), leading to a three-wave superposition and a 3D speckle field. Outside perturbations acting upon it induce intensity decorrelations, probed by a single mode fiber, a photon counter, and a correlator.

Taking into account not only the length of a chord $\Lambda_c$, but also the diffusive random walk inside the walls needed for the diffuse reflection process, we get that the total path length of a photon can be represented as a composite random variable:

$$\Lambda = n_c \Lambda_c + n_r c \tau_r, \tag{1}$$

where $n_r$ is the random number of reflections, $n_c = n_r + 1$ is the number of free-space chords through the cavity, and $\tau_r$ is a random variable that corresponds to the residency duration of the photon inside the wall at each reflection, also called reflection time. Note that the mean chord length within a cavity of volume $V_c$ and surface $\Sigma_c$ satisfies the well-known property $\overline{\Lambda}_c = 4V_c/\Sigma_c$ [12,16,17].

The phase $\phi$ of each wave is therefore associated with this optical length $\Lambda$, and can be rewritten introducing the aforementioned average chord length

$$\phi = \frac{2\pi \Lambda}{\lambda_0} = \frac{2\pi \overline{\Lambda}}{\lambda_0} + \frac{2\pi}{\lambda_0}(\Lambda - \overline{\Lambda}) = \overline{\phi} + k\tilde{\Lambda}, \tag{2}$$

where $\overline{\phi}$ is the average phase lag and $k\tilde{\Lambda} = k(\Lambda - \overline{\Lambda})$ is the phase shift specific to each path. In the experimental realization, $\overline{\Lambda}$ was measured $\sim$62 m, about eight orders of magnitude larger than the laser wavelength $\lambda_0$ [12]. Assuming that the standard deviation $\sigma_\Lambda$ is of the same order of magnitude than $\overline{\Lambda}$, we can safely assume that the standard deviation $\sigma_\phi$ of the phases is approximately eight orders of magnitude larger than $2\pi$, and that phases are thus uniformly distributed over $[0, 2\pi]$.

## III. MATHEMATICAL STRUCTURE OF THE RANDOM WAVE FIELD

A 3D field of random waves can be represented in a purely classical way as the real part $E(\mathbf{r}, t)$ of a complex vector field $\mathbf{E}(\mathbf{r}, t)$ in $\mathbb{C}^3$, constructed as the coherent sum of a large statistical set of elementary plane waves indexed by $\alpha \in \mathbb{N}$. Strictly speaking, the notion of a plane wave is only a local approximation. In our experiment, the field is probed locally by a single mode optical fiber and we can use this approximation for our calculation. Each elementary wave $\mathbf{E}_\alpha(\mathbf{r}, t)$ has a complex amplitude $a_\alpha e^{i\phi_\alpha}$ and carries a polarization represented by a complex unitary vector $\mathbf{d}_\alpha$ perpendicular to $\mathbf{k}_\alpha$, the wave vector. This elementary wave reads

$$\mathbf{E}_\alpha(\mathbf{r}, t) = a_\alpha e^{i\phi_\alpha} \mathbf{d}_\alpha \exp(i\mathbf{k}_\alpha \mathbf{r} - i\omega_\alpha t). \tag{3}$$

The key assumptions made by Berry and Dennis [18] to construct the 3D random wave model are as follows:

(1) Each elementary wave is a random object defined by three independent random variables: A wave vector $\mathbf{k}_\alpha$, a phase $\phi_\alpha$, and a polarization $\mathbf{d}_\alpha$.

(2) The random wave vector $\mathbf{k}_\alpha$ is uniformly distributed on the 3D sphere $|\mathbf{k}_\alpha| = k = 2\pi/\lambda_0$; the phase is uniformly distributed on $[0, 2\pi]$; the polarization state is uniformly distributed on the Poincaré sphere.

(3) The total field is a random sum of independent and identically distributed elementary waves.

The amplitudes $a_\alpha$ are normalization constants defined by the energy density of the field. Each statistical set of random waves $\{\mathbf{E}_\alpha(\mathbf{r}, t)_{1 \leqslant \alpha \leqslant N_{\mathcal{C}}}\}$ is referred to as a microscopic configuration of the field $\mathcal{C} = \{(a_\alpha, \mathbf{d}_\alpha, \phi_\alpha, \mathbf{k}_\alpha)_{1 \leqslant \alpha \leqslant N_{\mathcal{C}}}\}$, where $N_{\mathcal{C}}$ is the number of elementary waves:

$$\mathbf{E}(\mathbf{r}, t, \mathcal{C}) = \sum_{\alpha=1}^{N_{\mathcal{C}}} a_\alpha \mathbf{d}_\alpha e^{i\phi_\alpha} \exp(i\mathbf{k}_\alpha \mathbf{r} - i\omega t). \tag{4}$$

Each microscopic configuration $\mathcal{C}$ corresponds to a large number of degrees of freedom equal to $N_{\mathcal{C}}$ multiplied by the number of degrees of freedom of each field component. While there is no mathematical reason to set any particular upper bound to $N_{\mathcal{C}}$, the diffraction theory provides an estimate of this bound $\sim 10^{10}$ (see Appendix A).

Let us now consider a point $\mathbf{r}$ in the cavity and the time-averaged optical intensity $I(\mathbf{r}, \mathcal{C}) = c\epsilon_0 \langle E^2(\mathbf{r}, t, \mathcal{C}) \rangle_t$ of the real field $E(\mathbf{r}, t, \mathcal{C}) = \mathrm{Re}(\mathbf{E}(\mathbf{r}, t, \mathcal{C}))$ for a microscopic configuration $\mathcal{C}$. Using the complex vector formalism, it comes

$$\begin{aligned} I(\mathbf{r}, \mathcal{C}) &= c\epsilon_0 \, \langle \mathbf{E}(\mathbf{r}, t, \mathcal{C}) \mathbf{E}^*(\mathbf{r}, t, \mathcal{C}) \rangle_t \\ &= \frac{c\epsilon_0}{2} \left[ \sum_{\alpha=1}^{N_{\mathcal{C}}} a_\alpha \mathbf{d}_\alpha e^{i\phi_\alpha} e^{i\mathbf{k}_\alpha \mathbf{r}} \right] \left[ \sum_{\beta=1}^{N_{\mathcal{C}}} a_\beta \mathbf{d}_\beta e^{i\phi_\beta} e^{i\mathbf{k}_\beta \mathbf{r}} \right]^*. \end{aligned} \tag{5}$$

For a given configuration $\mathcal{C}$, the field complex amplitude obeys a zero-mean circular complex Gaussian density, and the speckle intensity is therefore exponentially distributed, as expected from a fully developed speckle. The field being fully polarized, it has a unity contrast, i.e., the standard deviation across space equals the spatial mean [19]:

$$\sigma_{I(\mathbf{r})}|\mathbf{r} = \langle (I(\mathbf{r}, \mathcal{C}) - I(\mathbf{r}', \mathcal{C}))^2 \rangle_{(\mathbf{r}, \mathbf{r}')} = \langle I(\mathbf{r}, \mathcal{C}) \rangle_{\mathbf{r}}. \tag{6}$$

These properties also apply in a dual way to the statistical distribution of the field and the intensity at a given point $\mathbf{r}$ across the microscopic configurations $\{\mathcal{C}\}$, with an exponential distribution of the intensity and a unity contrast, i.e., $\sigma_{I(\mathbf{r})|\mathcal{C}} = \langle I(\mathbf{r}) \rangle_{\mathcal{C}}$.

We also find that the mean intensity at a given point $\mathbf{r}$ over all microscopic configurations, $\langle I(\mathbf{r}) \rangle_{\mathcal{C}}$, does not depend on $\mathbf{r}$ and is equal to the mean intensity for a given configuration across space (showing an ergodicity property of our setup):

$$\langle I(\mathbf{r},\mathcal{C})\rangle_{\mathcal{C}} = \langle I(\mathbf{r},\mathcal{C})\rangle_{\mathbf{r}} = \overline{I}. \tag{7}$$

It can be factorized using the phase statistics in Eq. (2):

$$\begin{aligned}\langle I(\mathbf{r},\mathcal{C})\rangle_{\mathcal{C}} &= \frac{c\epsilon_0}{2}\sum_{1\leqslant\alpha,\beta\leqslant N_{\mathcal{C}}}\langle \mathbf{d}_\alpha \mathbf{d}_\beta^*\rangle_{\mathcal{C}}\langle e^{i(\mathbf{k}_\alpha-\mathbf{k}_\beta)\mathbf{r}}\rangle_{\mathcal{C}}\langle a_\alpha a_\beta e^{ik(\tilde{\Lambda}_\alpha-\tilde{\Lambda}_\beta)}\rangle_{\mathcal{C}}\\ &= \frac{c\epsilon_0 N_{\mathcal{C}}}{2}\langle a^2\rangle_{\mathcal{C}}.\end{aligned} \tag{8}$$

This simple result requires assumptions satisfied by the random field created in the 3D stochastic interferometer (see Appendix A).

## IV. INTERFEROMETRIC RESPONSE OF THE 3D SPECKLE FIELD

In our study, we only consider the perturbation of phase transport and disregard the first-order effects on the geometry of the elementary components of the field, i.e., the polarization $\mathbf{d}_\alpha$ and the direction $\mathbf{k}_\alpha/|\mathbf{k}_\alpha|$. We present a stochastic scalar perturbation theory that extends the classical two-arm interferometer theory to the case of a random number of isotropically distributed arms with random lengths. Furthermore, two metrics are commonly used to quantify the intrinsic sensitivity of a speckle pattern, namely, the spectral correlation function and the similarity [20]. However, those metrics apply to measured two-dimensional images of speckle patterns, while in the present case we will derive an intrinsic sensitivity from the more general Berry formalism. Moreover, this calculation is better suited to a temporal measurement made on a single speckle grain through a monomode fiber, for a frequency analysis, as we have done in Ref. [14].

## V. WAVELENGTH DERIVATIVE OF THE SPECKLE INTENSITY

Let us consider a microscopic configuration $\mathcal{C}$ of the random wave field, associated with the set of path lengths $\{\Lambda_\alpha\}_{\mathcal{C}}$, and compute the derivative $\partial_k I(\mathbf{r},\mathcal{C})$ of the intensity with respect to the wave number of the laser, $k = 2\pi/\lambda_0$

$$\begin{aligned}\partial_k I(\mathbf{r},\mathcal{C}) = \frac{c\epsilon_0}{2}\sum_{1\leqslant\alpha,\beta\leqslant N_{\mathcal{C}}}&[a_\alpha a_\beta \mathbf{d}_\alpha \mathbf{d}_\beta^* e^{ik(\tilde{\Lambda}_\alpha-\tilde{\Lambda}_\beta)}e^{i(\mathbf{k}_\alpha-\mathbf{k}_\beta)\mathbf{r}}\\ &\times[i(\tilde{\Lambda}_\alpha-\tilde{\Lambda}_\beta)+i(\mathbf{k}_\alpha-\mathbf{k}_\beta)\mathbf{r}/k]].\end{aligned} \tag{9}$$

Equation (9) can be interpreted with a thought experiment in which the wave vector of the input light changes from $k$ to $k+\delta k$. The first-order effect of that change on the intensity is given by $\delta I(\mathbf{r},\mathcal{C}) = \delta k \times \partial_k I(\mathbf{r},\mathcal{C})$. A comparison can be made here with a classical two-arm interferometer, in which the output intensity is not perturbed by $\delta k$ if its two arms have the same length. However, if the two arms have a known length difference $\Delta L$, the perturbation $\delta k$ can be computed from its deterministic consequence $\delta I$. Here instead, each configuration contains a large number of arms, and the interference results from an even larger number of arm pairs $(\alpha,\beta)$, which individually come with a random length difference $(\overline{\Lambda}_\alpha - \overline{\Lambda}_\beta)$. As a result, diagonal terms in Eq. (9) vanish, but nondiagonal terms do not, and they contribute to a nonzero random response characterized by $\partial_k I(\mathbf{r},\mathcal{C})$. However, because the nondiagonal terms $(\overline{\Lambda}_\alpha - \overline{\Lambda}_\beta)$ and $(\mathbf{k}_\alpha - \mathbf{k}_\beta)$ have zero expectation, the average response vanishes too, i.e., $\langle\partial_k I(\mathbf{r},\mathcal{C})\rangle_{\mathcal{C}} = 0$.

Therefore, our setup behaves as a stochastic interferometer, meaning that it responds to a given perturbation $\delta k$ by a stochastic response $\delta I(\mathbf{r},\mathcal{C}) = \delta k \times \partial_k I(\mathbf{r},\mathcal{C})$, which is a random variable of zero mean. In some way, this is what would happen with a classical two-arm interferometer in which the phase difference $\varphi_{1,2}$ is a uniform random variable in $[0, 2\pi]$. Since the sensitivity $\partial_\varphi I$ for a given phase difference is proportional to $\sin(\varphi_{1,2})$, it would then average out to zero over all configurations. Practically, no information can be extracted from using that interferometer with a single configuration, and no information can be obtained from the average response over a large number of configurations either.

The meaningful response is the quadratic response averaged over all microscopic configurations $\mathcal{C}$, as defined by

$$\langle\delta I(\mathbf{r},\mathcal{C})^2\rangle_{\mathcal{C}} = \delta k^2 \times \langle|\partial_k I(\mathbf{r},\mathcal{C})|^2\rangle_{\mathcal{C}}. \tag{10}$$

By introducing $M_{\alpha,\beta} = i(\tilde{\Lambda}_\alpha - \tilde{\Lambda}_\beta) + i(\mathbf{k}_\alpha - \mathbf{k}_\beta)\mathbf{r}/k$, square modulus of Eq. (9) reads

$$|\partial_k I(\mathbf{r},\mathcal{C})|^2 = \left[\frac{c\epsilon_0}{2}\right]^2 \sum_{1\leqslant\alpha,\beta,\gamma,\delta\leqslant N_{\mathcal{C}}} a_\alpha a_\beta a_\gamma a_\delta\, \mathbf{d}_\alpha \mathbf{d}_\beta^* \mathbf{d}_\gamma^* \mathbf{d}_\delta\, e^{ik[(\tilde{\Lambda}_\alpha-\tilde{\Lambda}_\beta)-(\tilde{\Lambda}_\gamma-\tilde{\Lambda}_\delta)]}e^{i(\mathbf{k}_\alpha-\mathbf{k}_\beta)\mathbf{r}}e^{-i(\mathbf{k}_\gamma-\mathbf{k}_\delta)\mathbf{r}}\, M_{\alpha,\beta}M_{\gamma,\delta}^*. \tag{11}$$

We now average this equation over all configurations, which amounts to considering the average for a random set of four waves $\langle\ldots\rangle_{(\alpha,\beta,\gamma,\delta)}$. The average over nondiagonal pairs of waves vanishes, i.e., $\langle\ldots\rangle_{(\alpha,\beta)\neq(\gamma,\delta)} = 0$, and only the diagonal terms corresponding to $(\alpha,\beta) = (\gamma,\delta)$ remain. For these diagonal pairs, the contribution of polarization vectors is 1 because they are unitary. Moreover, pairs $\alpha = \beta$ do not contribute as seen above in the analogy with equal arm interferometers.

We are left with computing the average $\langle a_\alpha^2 a_\beta^2 |M_{\alpha,\beta}|^2\rangle_{\alpha\neq\beta}$. Because variables $\tilde{\Lambda}$ and $\mathbf{k}$ are independent, and since $\langle\mathbf{k}_\alpha\mathbf{k}_\beta\rangle_{\alpha\neq\beta} = \langle\mathbf{k}_\alpha\rangle\langle\mathbf{k}_\beta\rangle_{\alpha\neq\beta} = 0$, it comes (see Appendix B for the detailed derivation)

$$\langle|\partial_k I(\mathbf{r},\mathcal{C})|^2\rangle_{\mathcal{C}} = N_{\mathcal{C}}(N_{\mathcal{C}}-1)\frac{c^2\epsilon_0^2}{2}[\langle a^2\rangle\langle a^2\tilde{\Lambda}^2\rangle - \langle a^2\tilde{\Lambda}\rangle^2]. \tag{12}$$

We observe that the statistics of the mean-square sensitivity are independent of **r**. From now on, we will omit the spatial variable **r** and consider a fixed arbitrary position $\mathbf{r} = 0$. Considering that $N_{\mathcal{C}} \gg 1$, we assume that the field amplitude $a$ and the path length $\Lambda$ are independent variables. Using the normalization of the intensity from Eq. (8), we find

$$\langle|\partial_k I|^2\rangle_{\mathcal{C}} = 2\langle I\rangle_{\mathcal{C}}^2\,\sigma_{\tilde{\Lambda}}^2, \tag{13}$$

where $\sigma_{\tilde{\Lambda}}^2 = \sigma_\Lambda^2$ is the variance of the path length distribution $\Lambda$. We know, however, from Eq. [21] that the derivative of the phase shift with the wave number (and therefore the path length $\Lambda$) has long-range correlations, and that the independence between $a$ and $\Lambda$ is a strong assumption. Appendix C details the exact calculation of the derivative of the intensity with the wave number in that context and gives an equivalent and consistent expression. From Eq. (1), we see that the exact value of the variance $\sigma_\Lambda^2$ requires the knowledge of variances $\sigma_{n_r}^2$, $\sigma_{\Lambda_c}^2$, and $\sigma_{\tau_r}^2$ of $n_r$, $\Lambda_c$, and $\tau_r$, respectively. We find

$$\sigma_\Lambda^2 = G^2\left[\overline{\Lambda}_c^2 + 2\sigma_{\Lambda_c}^2 + c^2\overline{\tau}_r^2 + 2c^2\sigma_{\tau_r}^2\right], \tag{14}$$

where $G = -1/\ln(\rho)$ is the cavity gain and $\rho$ is its albedo.

Practically in our experiments, the size of the cavity is large enough for the contribution of $c\,\tau_r$ to be negligible compared to $\Lambda_c$, and the variance of $\Lambda$ can be safely approximated accordingly, leading to $\sigma_\Lambda^2 \approx \sigma_{n_r\Lambda_c}^2 = G^2[\overline{\Lambda}_c^2 + 2\sigma_{\Lambda_c}^2]$. Using the ergodicity property $\langle I\rangle_{\mathcal{C}} = \overline{I}$ found in Eq. (7), it comes

$$\langle|\partial_k I|^2\rangle_{\mathcal{C}} = 2\,(G\overline{\Lambda}_c\overline{I})^2\left(1 + 2\,\frac{\sigma_{\Lambda_c}^2}{\overline{\Lambda}_c^2}\right). \tag{15}$$

In addition, the response to small variations of the wavelength is given by the derivative $\partial_\lambda I = -(k/\lambda)\,\partial_k I$:

$$\langle|\partial_\lambda I|^2\rangle_{\mathcal{C}} = 2(Gk\overline{\Lambda}_c\overline{I}/\lambda)^2\left(1 + 2\,\frac{\sigma_{\Lambda_c}^2}{\overline{\Lambda}_c^2}\right). \tag{16}$$

## VI. DERIVATIVE WITH RESPECT TO THE CAVITY GEOMETRY

Our experimental results show that the speckle intensity is sensitive to deformations of the cavity of the order of a few picometers under the conditions described in Eq. [14]. Similarly to the previous calculation, we show (see Appendix D) that the response to small variations of the cavity geometry (defined by a linear scaling parameter $\mu$) is given by

$$\langle|\partial_\mu I|^2\rangle_{\mathcal{C}} = 2(Gk\overline{\Lambda}_c\overline{I})^2\left(1 + 2\,\frac{\sigma_{\Lambda_c}^2}{\overline{\Lambda}_c^2}\right). \tag{17}$$

An analogous formula can also be obtained when considering a spatial variation of the refractive index $n = n_0 + \delta n$ of the cavity medium.

## VII. EXPERIMENTAL SENSITIVITY

In this section, we generalize the theoretical approach used in the previous sections to derive an expression for the sensitivity of the instrument with respect to any perturbation of the field (wavelength, cavity size, refractive index, or any quantity for which we can compute the partial derivative of the speckle intensity). We then numerically calculate the noise floor with experimental parameters, only to derive an expression in terms of a temporal decorrelation function.

Let us consider an experiment in which a measured intensity $I(\mathcal{C})$ is obtained from a microscopic configuration $\mathcal{C}$ of the speckle and a perturbation $\delta p$ applied to one parameter $p$ of the field. The variation of the intensity $\delta I = I(\mathcal{C}_{p+\delta p}) - I(\mathcal{C}_p)$ is given at first order by $\delta I \approx \delta p \times \partial_p I(\mathcal{C})$. $\delta I$ can therefore be seen as a random variable indexed on the set of field configurations $\{\mathcal{C}\}$. As discussed in the previous section, the meaningful quantity is the mean-square variation over $\{\mathcal{C}\}$, which reads $\langle\delta I^2\rangle_{\mathcal{C}} = \delta p^2 \times \langle|\partial_p I|^2\rangle_{\mathcal{C}}$.

The interferometric sensitivity $S_p$ to variations of the parameter $p$ can be generically defined as the ratio of the relative root-mean square (rms) intensity variations over the perturbation $|\delta p|$. The so-called rms sensitivity comes as

$$S_p = \frac{\sqrt{\langle\delta I_{(\delta p)}^2\rangle_{\mathcal{C}}}}{|\delta p|\,\overline{I}} = \frac{\sqrt{\langle|\partial_p I|^2\rangle_{\mathcal{C}}}}{\overline{I}}. \tag{18}$$

Using Eqs. (16) and (17), we obtain the rms sensitivities $S_\lambda$, $S_n$, or $S_\mu$ to changes of the wavelength, the index of refraction, or the scaling parameter $\mu$, respectively:

$$S_p = 2\sqrt{2}\pi\;G\,\overline{\Lambda}_c\,\lambda^{-\alpha}\sqrt{1 + 2\frac{\sigma_{\Lambda_c}^2}{\overline{\Lambda}_c^2}} \approx 10\,G\,\frac{\overline{\Lambda}_c}{\lambda^\alpha}, \tag{19}$$

where the exponent $\alpha$ is two for a wavelength perturbation and one for a deformation or a change in the refractive index. Practically, for the cavity used in previous publications [12]

$$\frac{\sqrt{\langle\delta I_{(\delta p)}^2\rangle_{\mathcal{C}}}}{\overline{I}} = \begin{cases} 1.4\times 10^{15}\,|\delta\lambda| \\ 9.4\times 10^{8}\,|\delta n| \\ 9.4\times 10^{8}\,|\delta\mu| \end{cases}. \tag{20}$$

Measuring the intensity subjected to a perturbation $\delta p$ with $n_{ph}$ photons, the previous equation gives the signal-to-noise ratio (SNR) associated with the intensity:

$$|\delta p|_{\min} = \frac{\lambda^\alpha}{2\pi G\overline{\Lambda}_c}\,\frac{\mathrm{SNR}}{\sqrt{n_{ph}}}. \tag{21}$$

As a result, with $n_{ph} = 10^6$ photons, e.g., with a 100 kHz counting rate during 10 s, and SNR $\geqslant 10$, we get $|\delta\lambda| \geqslant 1.1\times 10^{-17}$ m or $|\delta n| \geqslant 1.7\times 10^{-11}$, which are in good agreement with experimental realizations of other teams [22–24]. The detection limit $|\delta\lambda|_{\min} \approx 10^{-17}$m can be used to estimate the sensitivity to cavity deformations in our experiment. Using the relation $|\delta\overline{\Lambda}_c|/\overline{\Lambda}_c = |\delta\lambda|/\lambda$, we find a theoretical picometer-level sensitivity $|\delta\overline{\Lambda}_c|_{\min} \approx 10^{-12}$ m, in excellent agreement with our experimental measurements, about $2.7\times 10^{-12}$ m [12].

## VIII. CONCLUSION

In summary, we have shown that a superposition of coherent and random light waves provides a deterministic and quantifiable intensity response to external perturbations, such as changes in wavelength or refractive index. We derived the theoretical expected sensitivity of such instruments to external perturbations, including changes in the instrument geometry, that can access the picometer scale. This level of performance has been demonstrated experimentally using Lambertian cavities, which act as 3D stochastic interferometers.

Even if such intrinsically random fields have been extensively described after their first introduction by Berry [25–28], it has been so mostly within the framework of geometry and topology, with a focus on optical vortices and phase singularities [29], and very few direct experimental realizations within that context [30–32]. The essence of the interferometric response that we describe lies in the boundary conditions brought by Lambertian cavity walls, which introduce feedback as opposed to the boundless conditions used in Berry's theoretical description. Mostly used for high precision photometry [33,34] and occasionally coherent applications [22], we bring those cavities under the new light of the random wave model, bridging the gap between random light fields and interferometry.

3D stochastic interferometry has applications ranging from particle sizing, rheology, and marker-free protein study to even vibration sensing, and the theoretical foundation that we present in this study proves that it can be efficiently used to design high-sensitivity experiments. In addition, the field of random waves has been shown to raise deep fundamental questions in quantum mechanics or even superfluids, and this current study could shed new light on these phenomena, given its original experimental approach to the subject.

## ACKNOWLEDGMENTS

The authors are grateful to Prof. François Amblard, Prof. Bart van Tiggelen, Prof. John King, Prof. Joerg Enderlein, Prof. Philippe Poulin, Prof. Shankar Ghosh, and Prof. Rodney Ruoff for comments and discussions. This work has been supported by the Simons Foundation (Grants No. 601944 and No. 1027116, M.F.).

## DATA AVAILABILITY

The data that support the findings of this article are not publicly available. The data are available from the authors upon reasonable request.

## APPENDIX A

Equation (8) gives an expression for the mean intensity at a given point $\mathbf{r}$ over all microscopic configurations, $\langle I(\mathbf{r})\rangle_{\mathcal{C}}$. This result requires several assumptions listed in the main text. Here, we remind those assumptions and add their respective justifications:

(1) Each configuration $\mathcal{C}$ refers to a set of independent and identical random variables $\mathbf{E}_\alpha$ indexed by $\alpha$.

This issue of independence is directly related to the problem of the numbers of degrees of freedom of the optical field. From diffraction theory, the far field produced by the cavity walls can be fully represented with a finite set of wave vectors, the size of which is given in order of magnitude by $\Sigma_c/\lambda^2 \approx 10^{10}$. As a consequence, for a given configuration $\mathcal{C}$, the maximal number of degrees of freedom associated with the corresponding set of wave vectors $\{\mathbf{k}_\alpha\}_{\mathcal{C}}$ has the same order of magnitude. This order of magnitude does not include the degrees of freedom associated with the polarization and the phase, and thus far underestimates the number $N_{\mathcal{C}}$ of components that need to be considered in the sum. A further discussion would require to also analyze the nonclassical degrees of freedom associated with photon numbers. At any rate, the optical field has a finite number of degrees of freedom $N_{\mathcal{F}}$, and Eq. (5) provides a meaningful representation of the intensity field if it contains a number of components $N_{\mathcal{C}} \sim N_{\mathcal{F}}$. Under this condition, the random variables $\mathbf{E}_\alpha$ entering the representation of the field are independent.

(2) The two random vector variables $\mathbf{d}$ and $\mathbf{k}$ are mutually independent variables.

This assumption directly arises from the nature of the diffuse reflection process. As discussed above, this process is deterministic, but its complexity makes it appear as a random process that practically destroys any correlation between the polarization state and the wave vector for any reflected wave.

(3) $\mathbf{d}$ and $\mathbf{k}$ are independent of the phase and the real amplitude ($\phi$ and $a$, respectively).

For each path $\alpha$, the variables $\mathbf{d}_\alpha$ and $\mathbf{k}_\alpha$ are determined by the last diffuse reflection event. The phase and the real amplitude instead are determined by the whole stochastic structure of the path, and are therefore independent of $\mathbf{d}_\alpha$ and $\mathbf{k}_\alpha$. They are also mutually independent because of the circular uniformity of the phase.

(4) Because the polarization is isotropic, nondiagonal terms average to zero.

(5) $\mathbf{d}$ is a unitary variable in $\mathbb{C}^3$.

## APPENDIX B

As explained in the main text, only the terms such that $\alpha \neq \beta$ are left to compute in the sum from Eq. (11), which are as many as $N_{\mathcal{C}}(N_{\mathcal{C}}-1)$. The average $\langle a_\alpha^2 a_\beta^2 |M_{\alpha,\beta}|^2\rangle_{\alpha\neq\beta}$ can largely be simplified. Indeed, since $\langle \mathbf{k}_\alpha \mathbf{k}_\beta\rangle_{\alpha\neq\beta} = \langle \mathbf{k}_\alpha\rangle\langle \mathbf{k}_\beta\rangle_{\alpha\neq\beta} = 0$, all terms $k_{\alpha,\beta}^2$ and $k_\alpha k_\beta$ amount to 0 after averaging. Furthermore, since $\tilde{\Lambda}$ and $\mathbf{k}$ are independent, all terms $\tilde{\Lambda}_{\alpha,\beta} k_{\alpha,\beta}$ also average out to 0. Therefore, the remaining terms to average are $a_\alpha^2 a_\beta^2 |\tilde{\Lambda}_\alpha^2 + \tilde{\Lambda}_\beta^2 - 2\tilde{\Lambda}_\alpha\tilde{\Lambda}_\beta|$ . We then get the following simplification:

$$\left\langle a_\alpha^2 a_\beta^2 |M_{\alpha,\beta}|^2\right\rangle_{\alpha\neq\beta} = 2\langle a^2\rangle\langle a^2\tilde{\Lambda}^2\rangle - 2\langle a^2\tilde{\Lambda}\rangle\langle a^2\tilde{\Lambda}\rangle, \quad \text{(B1)}$$

and the final result

$$\langle|\partial_k I(\mathbf{r},\mathcal{C})|^2\rangle_{\mathcal{C}} = N_{\mathcal{C}}(N_{\mathcal{C}}-1)\frac{c^2\epsilon_0^2}{2}[\langle a^2\rangle\langle a^2\tilde{\Lambda}^2\rangle - \langle a^2\tilde{\Lambda}\rangle^2]. \quad \text{(B2)}$$

## APPENDIX C

Reference [21] provides an expression for the following joint probability distribution, under the assumption that the random wave field is a circular Gaussian process, a statement the Berry hypothesis relies on:

$$P(I,\phi',R) = \frac{I}{\pi Q a^2}\exp(-I)\times\exp\left[-\frac{I}{Qa^2}(\phi'-a^2)-\frac{I}{Qa^2}R^2\right], \quad \text{(C1)}$$

where $I$ is the intensity, $\phi'$ is the phase derivative with respect to the frequency $\omega$, $R$ is the derivative of the natural logarithm of the amplitude with respect to $\omega$, $a$ is the mean delay time, and $Q$ is a dimensionless parameter. Integrating $\phi'$ and using $(\log I)' = I'/I$, we find

$$P(I, I') = \frac{1}{\sqrt{\pi Q a^2 I}} \exp(-I) \times \exp\left(-\frac{(I')^2}{Q a^2 I}\right). \tag{C2}$$

This leads to

$$P(I') = \frac{1}{\sqrt{Q a^2}} \exp\left(-2\frac{|I'|}{\sqrt{Q a^2}}\right), \tag{C3}$$

hence

$$\langle (I')^2 \rangle = Q a^2. \tag{C4}$$

From Ref. [21], this value is proportional to the variance of delay times. Indeed, we have

$$Q a^2 = -C''(\Omega)_{\Omega=0} - a^2, \tag{C5}$$

where the term relative to the correlation matrix $C(\Omega)$ is proportional to $\langle t^2 \rangle$. Therefore, $\langle (I')^2 \rangle = Q a^2$ is proportional to the variance of the delay time, which is consistent with the expression found in Eq. (13), where we find a proportionality to the variance of the path length.

## APPENDIX D

To theoretically derive intensity variations from deformations of the cavity is a difficult problem, because there is no mathematical representation of cavity deformations that would be simple enough to derive their effect on the speckle intensity. To circumvent this difficult problem and provide a simplistic but heuristic solution, we use a simple perturbation approach. To represent small cavity deformations that cause a change $\delta \overline{\Lambda}_c$ of the mean-chord invariant $\overline{\Lambda}_c = 4V_c/\Sigma_c$, we introduce the dimensionless scaling parameter $\mu = 1 + \delta\mu$ and substitute the random variable $\Lambda$ that represents random chords by $\mu\Lambda$. In this approach, we implicitly assume that the sole effect of $\delta\mu$ is to induce a phase shift on the elementary fields $\mathbf{E}_\alpha$, without changing the geometry of these plane waves.

This strategy relies on the notion that the physics does not change if the scale that measures the wavelength and the scale that measures the geometry of the cavity down to its smallest microscopic details are changed by the same factor. By the same token, inasmuch a cavity deformation can be considered to be equivalent to a dilation by the factor $\mu = 1 + \delta\mu$, its effect will be equivalent to dilation of the wavelength by the factor $1/\mu$, i.e., changing the wave vector by the factor $\mu$.

With this dimensionless dilation scaling parameter $\mu$, the phase factor $ik(\tilde{\Lambda}_\alpha - \tilde{\Lambda}_\beta)$ is substituted by $ik\mu(\tilde{\Lambda}_\alpha - \tilde{\Lambda}_\beta)$, and the speckle intensity can be derived with respect to $\mu$. A so-called scale-derivative $\partial_\mu I = k\partial_k I$ is obtained, which reads

$$\langle |\partial_\mu I|^2 \rangle_C = 2(G k \overline{\Lambda}_c \overline{I})^2 \left(1 + 2\frac{\sigma^2_{\Lambda_c}}{\overline{\Lambda}_c^2}\right). \tag{D1}$$

[1] Y. Wan, X. Fan, and Z. He, Review on speckle-based spectrum analyzer, Photonic Sens. **11**, 187 (2021).

[2] N. K. Metzger, R. Spesyvtsev, G. D. Bruce, B. Miller, G. T. Maker, G. Malcolm, M. Mazilu, and K. Dholakia, Harnessing speckle for a sub-femtometre resolved broadband wavemeter and laser stabilization, Nat. Commun. **8**, 15610 (2017).

[3] V. Tran, S. K. Sahoo, D. Wang, and C. Dang, Utilizing multiple scattering effect for highly sensitive optical refractive index sensing, Sens. Actuators A: Phys. **301**, 111776 (2020).

[4] A. Vijayakumar, D. Jayavel, M. Muthaiah, S. Bhattacharya, and J. Rosen, Implementation of a speckle-correlation-based optical lever with extended dynamic range, Appl. Opt. **58**, 5982 (2019).

[5] B. J. Berne and R. Pecora, *Dynamic Light Scattering: With Applications to Chemistry, Biology, and Physics* (John Wiley & Sons, New York, 1976).

[6] G. Maret and P. E. Wolf, Multiple light scattering from disordered media. The effect of Brownian motion of scatterers, Z. Phys. B Condens. Matter **65**, 409 (1987).

[7] D. J. Pine, D. A. Weitz, P. M. Chaikin, and E. Herbolzheimer, Diffusing wave spectroscopy, Phys. Rev. Lett. **60**, 1134 (1988).

[8] F. C. MacKintosh and S. John, Diffusing-wave spectroscopy and multiple scattering of light in correlated random media, Phys. Rev. B **40**, 2383 (1989).

[9] G. Graciani, L. Le Goff, and F. Amblard, Protein conformational dynamics probed correlation spectroscopy of multiply scattered light, Biophys. J. **118**, 136a (2020).

[10] P. Zakharov and F. Scheffold, Advances in dynamic light scattering techniques, in *Light Scattering Reviews 4*, edited by A. A. Kokhanovsky (Springer, Berlin, Heidelberg, 2009), pp. 433–467.

[11] H. S. Kim, N. Şenbil, C. Zhang, F. Scheffold, and T. G. Mason, Diffusing wave microrheology of highly scattering concentrated monodisperse emulsions, Proc. Natl. Acad. Sci. USA **116**, 7766 (2019).

[12] G. Graciani, M. Filoche, and F. Amblard, 3D stochastic interferometer detects picometer deformations and minute dielectric fluctuations of its optical volume, Commun. Phys. **5**, 239 (2022).

[13] G. Graciani and F. Amblard, Random dynamic interferometer: Cavity amplified speckle spectroscopy using a highly symmetric coherent field created inside a closed Lambertian optical cavity, in *Applied Optical Metrology III*, edited by E. Novak and J. D. Trolinger (International Society for Optics and Photonics, SPIE, San Diego, 2019), Vol. 11102, p. 111020N.

[14] G. Graciani, J. T. King, and F. Amblard, Cavity-amplified scattering spectroscopy reveals the dynamics of proteins and nanoparticles in quasi-transparent and miniature samples, ACS Nano **16**, 16796 (2022).

[15] G. Graciani, J. T. King, and F. Amblard, Marker-free protein study by amplified light scattering, in *SPIE Advanced Biophotonics Conference*, edited by E. Chung, K.-H. Jeong, C. Joo, W. Jung, H.-W. Kang, C.-S. Kim, C. Kim, P. Kim, and H. Yoo (International Society for Optics and Photonics, SPIE, Busan, Republic of Korea, 2022), Vol. 12159, p. 1215908.

[16] R. F. S. Blanco, An invariance property of diffusive random walks, Europhys. Lett. **61**, 168 (2003).

[17] R. Savo, R. Pierrat, U. Najar, R. Carminati, S. Rotter, and S. Gigan, Observation of mean path length invariance in light-scattering media, Science **358**, 765 (2017).

[18] M. V. Berry and M. R. Dennis, Phase singularities in isotropic random waves, Proc. R. Soc. Lond. A **456**, 2059 (2000).

[19] J. Uozumi, K. Uno, and T. Asakura, Statistics of Gaussian speckles with enhanced fluctuations, Opt. Rev. **2**, 174 (1995).

[20] M. Facchin, S. N. Khan, K. Dholakia, and G. D. Bruce, Determining intrinsic sensitivity and the role of multiple scattering in speckle metrology, Nat. Rev. Phys. **6**, 500 (2024).

[21] B. van Tiggelen, P. Sebbah, M. Stoytchev, and A. Genack, Delay-time statistics for diffuse waves, Phys. Rev. E **59**, 7166 (1999).

[22] M. Facchin, K. Dholakia, and G. D. Bruce, Wavelength sensitivity of the speckle patterns produced by an integrating sphere, J. Phys. Photonics **3**, 035005 (2021).

[23] M. Facchin, G. D. Bruce, and K. Dholakia, Measurement of variations in gas refractive index with $10^{-9}$ resolution using laser speckle, ACS Photonics **9**, 830 (2022).

[24] M. Facchin, G. D. Bruce, and K. Dholakia, Measuring picometre-level displacements using speckle patterns produced by an integrating sphere, Sci. Rep. **13**, 14607 (2023).

[25] G. Iooss, R. H. Helleman, and R. Stora, Chaotic behaviour of deterministic systems, *Les Houches Summer School Proceedings* (North-Holland, Amsterdam, 1983), Vol. 36.

[26] P. O'Connor, J. Gehlen, and E. J. Heller, Properties of random superpositions of plane waves, Phys. Rev. Lett. **58**, 1296 (1987).

[27] M. Berry and M. Dennis, Polarization singularities in isotropic random vector waves, Proc. R. Soc. Lond. A **457**, 141 (2001).

[28] L. De Angelis, F. Alpeggiani, A. Di Falco, and L. Kuipers, Spatial distribution of phase singularities in optical random vector waves, Phys. Rev. Lett. **117**, 093901 (2016).

[29] N. Shvartsman and I. Freund, Vortices in random wave fields: Nearest neighbor anticorrelations, Phys. Rev. Lett. **72**, 1008 (1994).

[30] N. Jhajj, I. Larkin, E. W. Rosenthal, S. Zahedpour, J. K. Wahlstrand, and H. M. Milchberg, Spatiotemporal optical vortices, Phys. Rev. X **6**, 031037 (2016).

[31] F. Eilenberger, K. Prater, S. Minardi, R. Geiss, U. Röpke, J. Kobelke, K. Schuster, H. Bartelt, S. Nolte, A. Tünnermann, and T. Pertsch, Observation of discrete, vortex light bullets, Phys. Rev. X **3**, 041031 (2013).

[32] W. Wang, S. G. Hanson, Y. Miyamoto, and M. Takeda, Experimental investigation of local properties and statistics of optical vortices in random wave fields, Phys. Rev. Lett. **94**, 103902 (2005).

[33] I. Khaoua, G. Graciani, A. Kim, and F. Amblard, Stochastic light concentration from 3D to 2D reveals ultraweak chemi- and bioluminescence, Sci. Rep. **11**, 10050 (2021).

[34] I. Khaoua, G. Graciani, A. Kim, and F. Amblard, Detectivity optimization to measure ultraweak light fluxes using an EM-CCD as binary photon counter array, Sci. Rep. **11**, 3530 (2021).